# XprospeCT: CT Volume Generation from Paired X-Rays

Benjamin Paulson, Joshua Goldshteyn, Sydney Balboni, John Cisler, Andrew Crisler, Natalia Bukowski, Julia Kalish, Theodore Colwell

Department of Electrical Engineering and Computer Science

Milwaukee School of Engineering

1025 N Broadway St, Milwaukee, WI 53202

(paulsonb, goldshteynj, balbonis, cislerj, crislera, bukowskin, kalishj, colwellt)
@msoe.edu

## Abstract

Computed tomography (CT) is a beneficial imaging tool for diagnostic purposes. CT scans provide detailed information concerning the internal anatomic structures of a patient, but present higher radiation dose and costs compared to X-ray imaging. In this paper, we build on previous research to convert orthogonal X-ray images into simulated CT volumes by exploring larger datasets and various model structures. Significant model variations include UNet architectures, custom connections, activation functions, loss functions, optimizers, and a novel back projection approach.

# 1 Introduction

Computed tomography (CT) allows for detailed views of anatomic structures by reconstructing hundreds of X-ray images taken from a full rotation around a patient's body at various angles to produce three-dimensional volumes. The information acquired by a CT scan is beneficial to the diagnostic process as an exploratory analysis tool but is resource-demanding and requires high levels of ionizing radiation exposure with an average effective radiation dose of 7 mSv per chest CT [4]. In contrast, two-dimensional X-ray imaging is a widely used medical imaging modality as it is resource-efficient and results in a lower average radiation dose of 0.1 mSv per chest X-ray [4]. Although more efficient, X-rays lack the detailed spatial information obtained from a three-dimensional imaging modality.

Many healthcare protocols involve an X-ray scan for preliminary information gathering, allowing healthcare workers to determine the need for risking higher radiation exposure and expending more resources with additional imaging modalities. Although X-ray scans provide a clear view of bone structures, soft tissues are less defined. Machine learning methods have shown to be an effective way to extrapolate the two-dimensional information from X-rays into three-dimensional space, remedying the shortcomings of traditional X-rays while equipping professionals with additional information for making well informed decisions.

We propose to improve upon past research by making a model capable of generating simulated CT scans from X-ray image inputs with high detail. In the recent work, "X2CT-GAN: Reconstructing CT from Biplanar X-Rays with Generative Adversarial Networks", Ying *et al.* researched the simulation of CT scans from orthogonal X-ray views [1]. This paper's approach included a Generative Adversarial Network (GAN) framework with a specialized generator. We hypothesize that using a larger dataset of CT exams and testing several network architectures will allow for improved reconstruction results.

In this work, we develop deep-learning models for reconstructing CT scans from paired lateral and anterior-posterior X-rays to generate an accurate volume for structures in the chest region. These reconstructed CT scans may allow for the preliminary detection of medical abnormalities, enabling a patient's care team to determine the need for additional resource-intensive imaging methods.

# 2 Data

The models in this study depend on two distinct datasets: the Rad-ChestCT dataset [2] and the CheXpert dataset [3]. The former dataset was obtained from the Center for Virtual Imaging Trials at Duke University and consists of 35,747 chest CT scans acquired from 19,661 adult patients. Access was granted to 10% of the data, comprising 3,630 chest scans. This dataset was developed by Rachel Draelos, an MD/PhD student, with the objective of facilitating machine learning models focused on CT scans. The CheXpert dataset is an open-source dataset that comprises of 224,316 chest radiographs taken from 65,240 patients. For training the style transformation model, a total of 2,390 paired orthogonal X-rays were used. To be used as the ground truth in the reconstruction model, the CT images were normalized by scaling the volumes from -1000, 1000 HU to between 0 and 1. Next, these volumes were resized to either 128×128×128 or 64×64×64 to be inputs for the reconstruction model. The dataset was split, with 80% partitioned for training and 20% for validation. Furthermore, 30 test images were extracted and used to compare models as shown in the Experiments Table (Table 1).

# 3 Models

Several experiments were performed using simulated X-ray images as the input for the reconstruction model, which takes a set of paired X-rays and yields a simulated CT volume. The creation of simulated X-rays required style transfer models – one for each X-ray view – which created paired data from the Rad-chest CT [2] and CheXpert [3] datasets.

## 3.1 Style Transfer General Adversarial Network

Obtaining paired data from the separate CheXpert and Rad-ChestCT datasets presented a challenge. Past research into a model capable of learning the mapping between two unpaired datasets resulted in CycleGAN [5]. We used CycleGAN as a style transformation model to obtain paired data for training. The style transformation model was trained on X-rays from the CheXpert dataset and mean CT scans that were originally a part of the Rad-ChestCT dataset. The style transformation model was employed to learn the style of a real X-ray that can then be transferred to mean CT scans in the process of generating paired data.

The CT scans of the Rad-ChestCT dataset were averaged along the coronal and sagittal axis to obtain two-dimensional CT scans in both the anterior-posterior and lateral views. The mean CT scans were inputs to the style transfer general adversarial network which then stylized the mean CT scans to have a closer resemblance to real X-rays. This process resulted in paired simulated X-rays and CT scans that were used to train the reconstruction model in some experiments.

Changes were made to the CycleGAN model to better process the style transformation that is required to go between mean CT scans and X-rays. The first major change included significantly decreasing the learning rate, as the level of the style transform amounted to a non-linear contrast transformation as opposed to transitioning between two different styles. This was paired with the use of an exponential moving average, which reset the weights to their moving average to increase the generalization of the model. This also allowed for finer level updates between epochs.

On top of these changes, the Nadam optimizer was used instead of Adam due to the better level of convergence that it achieved. The Nadam and Adam optimizers work similarly, except for incorporating Nesterov momentum which performs updates to the gradient based on the projected update, as opposed to the current value of each parameter. This allows for a lower level of overshooting once a local minimum has been found and has resulted in better styled X-rays than those obtained using pure Adam.

Another change was made to increase the Lambda value to 20. The Lambda variable describes the cycle consistency loss which affects how accurate the style reconstruction is by comparing the original input when going both forward and backward through the CycleGAN training. This was originally set to 10 for most uses, but it was discovered that the pixel alignment improved the style transfer when using a higher value by testing different values. However, if the value was too high (greater than 25), the style transfer ended up becoming very weak and the differences between the input and output became insufficient. With a lambda of 20, a style transfer occurred that included significant X-ray characteristics (such as less contrast of different depth regions as compared to mean CT scans) while maintaining a similar shape to that of the original.

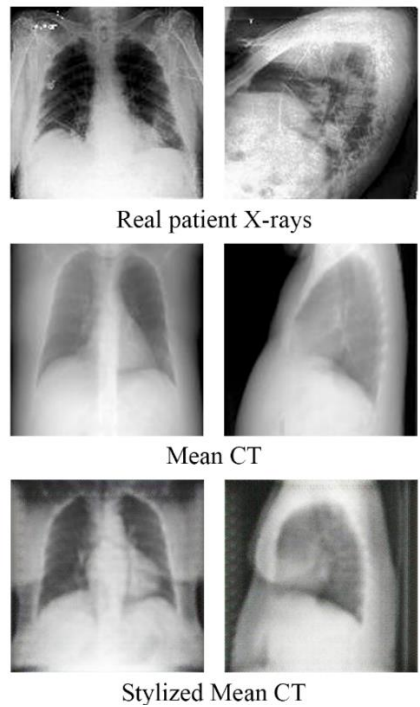

*Figure 1: Examples of real Patient Coronal and Sagittal View X-rays from the CheXpert Dataset, Mean Coronal and Sagittal View X-rays, and Stylized Coronal and Sagittal View X-rays*

## 3.2 Reconstruction Model

The reconstruction model is responsible for the transformation of two orthogonal X-rays to a simulated CT volume. Our reconstruction model is inspired by the "X2CT-CNN+B" reconstruction model proposed in the paper "X2CT-GAN: Reconstructing CT from Biplanar X-Rays with Generative Adversarial Networks" [1], comprised of two UNet encoder-decoder architectures for corresponding simulated X-ray inputs. Following the decoders, a fusion network with custom skip connections interweaved from the orthogonal decoder architectures iteratively extracts their convolved features into the yielded simulated CT volume. Modifications made to the initial "X2CT-CNN+B" reconstruction model – as well as associated result discussions – are compiled in "Section 4: Experiments & Results". Featured experimental adjustments ranged from architectural to hyperparameter changes. Contrasting previous work, the discussed reconstruction models were trained using paired images of real, as opposed to averaged, orthogonal X-Ray images and associated CT scans in some experiments.

Stemming from challenges faced regarding bit limitations of Keras Tensors (see 6.3) [7], the initial architecture (Figure 2) was modified from the "X2CT-CNN+B" for 128×128-pixel images rather than 256×256-pixel images, resulting in significant memory savings due to CT labels consequently reduced to 128×128×128 voxel scans. The UNet architecture was chosen due to its ability to preserve features – via skip connections – from input images regardless of the convolutional changes that occur throughout the encoding portion of the network. This preservation through skip connections was important due to the emphasis on small details in medical-imaging applications.

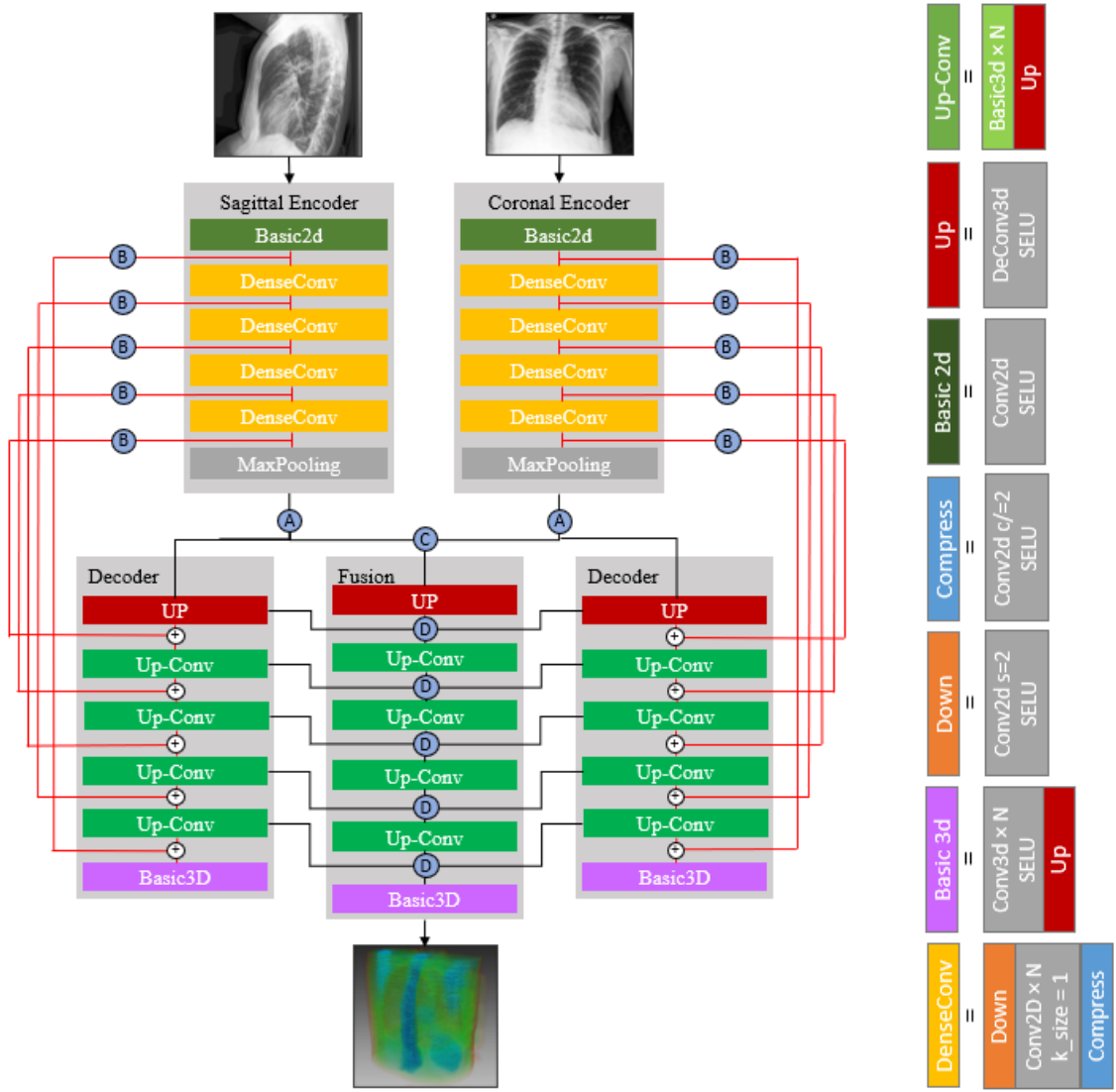

*Figure 22: Dense-512 Reconstruction Model with Model-Layer Legend*

## 3.3 Reconstruction Model Connections

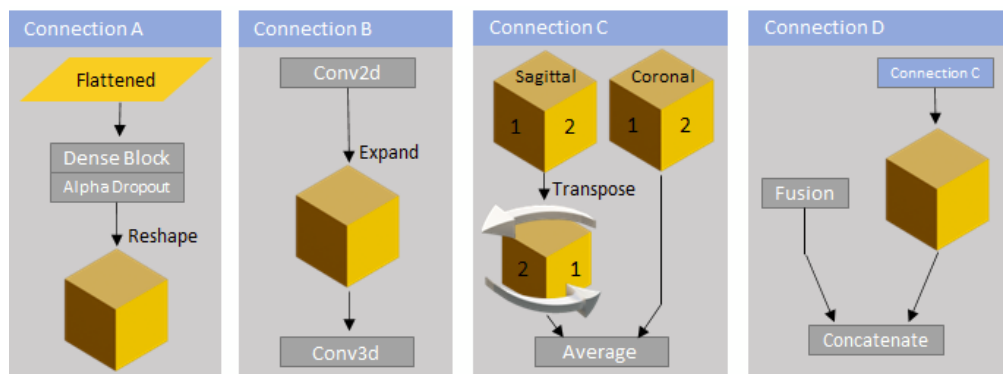

*Figure 33: Architectural View of Connections*

The connections that link the individual components of the reconstruction model are divided as Connection-A, Connection-B, Connection-C, and Connection-D. The placement of these connections within the overall architecture did not change across experiments; however, if the connection was altered, the internal definition was changed without changing the external I/O of each connection. This was done to maintain the capability of the overall structure while testing smaller model changes.

**Connection-A** reshapes the input tensor to a uniform three-dimensional output regardless of the input dimensionality. Following the convention for tensors in Keras, this is represented by a five-dimensional tensor formatted in order of following index: batch size, rows, columns, depth, and channels. Further discussion about the final Connection-A architecture is provided in section 4.3.

**Connection-B** augments the skip connections of each UNet architecture present throughout the reconstruction model. Because the decoding portion of the UNet consists of five-dimensional tensors – but the skip connections from correlating encoding layers are four-dimensional – the skip connections must also account for transforming two-dimensional image data to a three-dimensional representation to be concatenated with the decoded layers' outputs. The presence of Conv2D and Conv3D layers within the connection respectively match the number of channels and transform to an expanded five-dimensional tensor.

**Connection-C** transposes the sagittal and coronal voxel representations such that the sagittal view is appropriately orthogonal to the correlating front-facing coronal voxel representation. The result of this permutation is then averaged to produce a coherent voxel representation for input into the decoder layers of the fusion network.

**Connection-D** uses Connection-C yet accepts an additional input from the previous decoding layer of the fusion network to be concatenated with the averaged output of Connection-C.

# 4 Experiments & Results

Below is a brief, tabular overview of the experiments discussed in this section.

| Model Name | Input X-ray Images | Training Loss function | Optimizer | Test Accuracy MSE (±0.001) | Example Axial Image |
|---|---|---|---|---|---|
| **64-Dense** | Mean CT | MSE | Adam | 0.0834 | 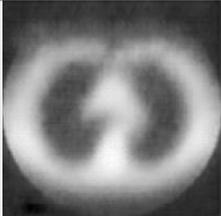 |
| **64-Dense-Styled-Inputs** | Stylized CT | MSE | Adam | 0.0751 | 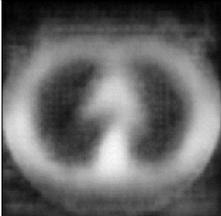 |
| **512-Dense** | Mean CT | MSE | Adam | 0.0819 | 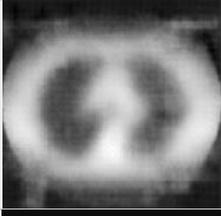 |
| **512-Dense MAE** | Mean CT | MAE | Nadam | 0.0970 | 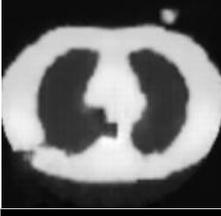 |
| **512-ProjInj** | Mean CT | MAE | Nadam | 0.0980 | 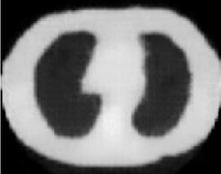 |

*Table 1: Model names, input x-ray images, training loss function, optimizer, test accuracy, and example axial images.*

**Note:** In every experiment, the SELU activation function is used due to its self-normalizing properties. This allows for a faster and more accurate level of training than using ReLU layers followed by batch normalization.

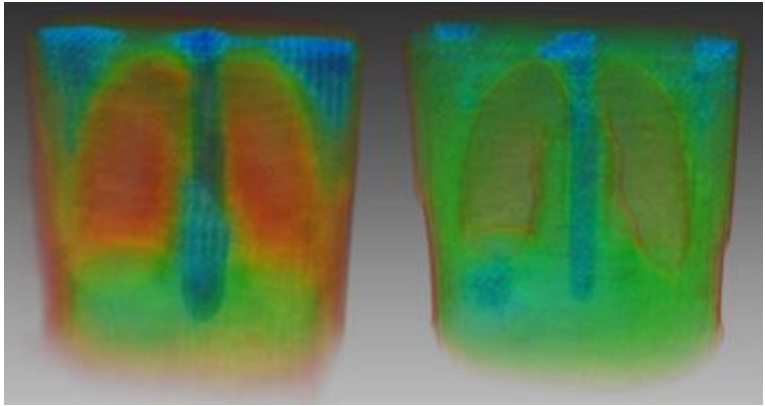

*Figure 4: 3D Volume Renderings of predictions (64-Dense-Styled-Inputs [Left], 512-ProjInj [Right]). Features including the lungs and spine can be seen. All models failed to synthesize dense bones, as shown by the lack of ribs.*

## 4.1 Input X-Ray Dimensions

Several experiments were performed with various input dimensions (not included in experiment table). An analysis was performed on image pixel sizes of 128×128 and 64×64 – outputs were kept at similar dimensionality matching the input dimension height and length.

The 256×256 input images were unable to train due to TensorFlow's tensor size limit (Section 5.2). The 128×128 with a "64-Dense" architecture produced results comparable to "X2CT-CNN+B" [1]. Additional experiments were performed with a 64×64 input, further decreasing the required memory and training time for the stylized CheXpert dataset; however, the decreased amount of encoding/decoding layers resulted in significant artifacts for predicted CT-Scans due to the lack of model structure to recognize specific features. Furthermore, the reduced resolution of the input X-rays restricts the amount of data the model must recognize to identify key features and patterns from the mean CT scans.

## 4.2 Training Loss Function & Optimizer

Initial experiments with choice of training loss function and optimizer followed guidelines set by existing work [1]; these guidelines are Mean-Squared Error Loss and Adam Adaptive Optimizer [9]. Results with these guidelines produced CT scans with noise surrounding key X-Ray features: lungs, spine, and rib cage.

While noisy output may indicate a deep-learning model deficiency in adequate capacity for the application, variations around key features may indicate an inappropriate loss function. Mean-Squared Error as the training loss function lacks spatial awareness – a key factor in outputting noise-minimized CT scans. This resulted in multiple experiments with varying loss functions, resulting in Mean Absolute Error (MAE) being identified as the best candidate for effectively reducing the blurriness of the output CT scan data. While usage of MAE as a loss function reduces the precision of output features, this model is intended to be used for preliminary imaging purposes rather than full medical diagnoses.

Learning-rate optimization was initially performed with Adaptive Moment Estimation (Adam) to use the low memory requirements and capability to effectively optimize the training process of a large dataset [9]. However, Adam has significant difficulties escaping local minima whereas Nadam incorporates Nesterov-Accelerated momentum to minimize these convergence issues. Nadam often produces models with better generalization capabilities in less training time [10]. Experiments with Nadam as the optimizer, along with the benefits of MAE as the associated training loss function produced the clearest results throughout the reconstruction model experiments.

Ultimately, MAE and Nadam were chosen due to reduced blurriness within the CT scan output. However, this decreased blurriness does not mean the CT scan is less accurate due to the inherent quality of MAE and Nadam that tend to perform very similarly to the more commonly used MSE and Adam in other scenarios [10]. The MAE loss function allowed for predicting model losses linearly as opposed to squaring the error and treating each degree of error with the same amount of importance. Using Nadam allowed for the use of Nesterov acceleration with Adam, which updated the momentum weights in response to the future location as opposed to the current location (potentially preventing overshooting in local minima).

## 4.3 Connection-A Architecture

The first two experiments involved differing Connection-A architectures involving the number of dense-layer neurons at the apex of both the sagittal and coronal UNet architectures. "64-Dense" and "512-Dense" distinguish these experiments as the number of neurons within the dense block. Minimal visual and quantitative differences were detected between these changes; however, "512-Dense" produced CT-Scans more aware of features outside the immediate intrathoracic cavity such as shoulder blades and upper and lower vertebrae.

## 4.4 Back Projection Injection

Back projection injection is the largest architectural change to the previous alterations of "X2CT-CNN+B" mentioned above. This novel approach includes a simplified version of back projection with the orthogonal X-ray images. It produces CT scans with increased visibility of granular yet key features such as the vertebrae. This approach was considered because back projection is the result of a calculation, preserving spatial data, rather than requiring additional data collection and is similar to conventional CT reconstruction methods.

# 5 Challenges

## 5.1 Memory Limitations of Single-GPU Training

During the creation of the reconstruction model outlined in (Figure 2), we came to the realization that the quantity of layers used in combination with the image size were consuming a lot of memory during run-time. The biggest cause of our memory was the final size of our output: 128×128×128 pixels. Many of the intermittent models tried to use high pixel inputs and outputs to increase accuracy of the result but failed due to reaching the memory limit. This restricted the maximum quality of the outputs produced. The final reconstruction model consumed around 25 GiB of memory and the Projection Injection consumed around 32 GiB of memory.

## 5.2 TensorFlow's Max Limit on Parameters

Another issue encountered during testing of model architectures was the TensorFlow library limitation on max parameters in a layer. TensorFlow limits its parameters per layer at $2^{32}$ parameters. This also contributed to our max possible output size being 128×128×128 pixels, again restricting the max quality of our results. We ended up working within Keras' layer constraint, but future work could look at solving this problem, potentially with PyTorch (Section 6.1) to create more detailed results.

# 6 Future Work

## 6.1 Switch to PyTorch to Solve Max Parameters Constraint

As mentioned in the Challenges, TensorFlow's Keras library has a limited number of parameters allowed per layer. This resulted in our reconstruction model being limited to an output of 128×128×128 for the simulated CT scan. To achieve more defined simulated CT scans, the model output layer needs more parameters. The best alternative we identified was switching to PyTorch's libraries [10], which would require a large code change.

## 6.2 Transformer Architecture

Currently, all models tested directly input the coronal and sagittal X-Ray images into the reconstruction model's encoders, as shown in (Figure 2). However, past research [6] has shown that a multi-headed attention layer before the encoder could lead to increased quality in the final picture.

### 6.3 Multi-GPU Training

The implementation of the reconstruction model uses a single GPU to train. However, this was relatively slow, limiting the number of epochs that models could train for. Using multiple GPUs may reduce training time and allow for the model to get better results by training across more epochs.

### 6.4 Use Medical Label AI to Test Accuracy of Output

Other research [3] has attempted to label CT scans from the Rad-ChestCT data set. To verify the validity of the output from the models, it could be beneficial to match labels found in both the Rad-ChestCT and CheXpert datasets and then train the model on the Rad-ChestCT data and verify the accuracy using CheXpert as the validation set. This would ensure that the model is reconstructing the CT scans properly instead of losing information in the process.

### 6.5 Train on Larger-scale X-Ray Images

To speed up training times and reduce run-time memory consumption, it was necessary to reduce the input X-Rays to 128×128. However, this traded the maximum quality of the resulting CT scans. To improve the reconstruction model, scaling up the input images to at least 256×256 or potentially higher may be useful. This will increase the train time and memory consumption, but with other future directions mentioned above, these effects could be reduced.

### 6.6 Partner with Medical Experts to Evaluate Usability

In the future, it would be beneficial to validate this model with medical experts to determine its relevance in clinical settings. Experts could also identify areas of improvement, identifying possible future research directions.

## 7 Conclusion

This work explores various techniques to reconstruct CT scans from paired orthogonal X-rays. Usage of the reconstruction model was primarily used to facilitate the generation of simulated CT volumes, along with several experiments performed using simulated X-ray images created by the style transformer model. While the stylized model (table 1, row 2) has the best mean squared error value, we consider the back projection model (table 1, row 5) to provide more accurate results for granular features within the final CT scan compared to other models that were experimented with. We hope to further our research by using multiple GPUs, adding attention to our reconstruction model, improving input X-Ray image size, and look at evaluating our model, both by using CT labeling models as well as receiving feedback from medical experts. With further research, we hope to create a model that clinicians can use to improve the diagnostic process.